\documentclass[sigconf,screen]{acmart}

\AtBeginDocument{%
  }

\copyrightyear{2025}
\acmYear{2025}
\setcopyright{cc}
\setcctype{by-sa}
\acmConference[SIGIR-AP 2025]{Proceedings of the 2025 Annual International ACM SIGIR Conference on Research and Development in Information Retrieval in the Asia Pacific Region}{December 7--10, 2025}{Xi'an, China}
\acmBooktitle{Proceedings of the 2025 Annual International ACM SIGIR Conference on Research and Development in Information Retrieval in the Asia Pacific Region (SIGIR-AP 2025), December 7--10, 2025, Xi'an, China}\acmDOI{10.1145/3767695.3769509}
\acmISBN{979-8-4007-2218-9/2025/12}

\usepackage{algorithmic}
\usepackage{textcomp}
\usepackage{multirow}
\usepackage{graphicx} 
\usepackage{subcaption} 
\usepackage{booktabs}
\usepackage{ragged2e}
\usepackage[normalem]{ulem}
\usepackage{xspace}
\usepackage{caption} 
\usepackage{array}
\usepackage{makecell}
\usepackage{bookmark}
\usepackage[table,xcdraw]{xcolor}
\usepackage{colortbl}
\usepackage{soul}
\usepackage{adjustbox}
\usepackage{enumitem}

\usepackage{xcolor}
\usepackage[most]{tcolorbox}
\usepackage{tabularx} 

\usepackage{amsmath} 
\usetikzlibrary{positioning, shapes.geometric, calc}
\usetikzlibrary{arrows.meta, bending, shapes.misc}

\usepackage{enumitem}
\setlist[itemize]{leftmargin=2em}
\setlist[enumerate]{leftmargin=2em}
\setlist[description]{leftmargin=2em}

\newcommand{\eg}{{e.g., }}
\newcommand{\ie}{{i.e., }}

\newcommand{\ds}[1]{{\textcolor{purple}{[DS: #1]}}}
\newcommand{\dah}[1]{{\textcolor{teal}{[DH: #1]}}}

\definecolor{specialblue}{HTML}{018786}
\newcommand{\sq}[1]{{\textcolor{specialblue}{[Sq: #1]}}}
\definecolor{specialpurple}{HTML}{6A1B9A}
\newcommand{\reviewer}[1]{{\textcolor{red}{[Reviewer: #1]}}}

\renewcommand{\reviewer}[1]{}
\renewcommand{\sq}[1]{}
\renewcommand{\ds}[1]{}
\renewcommand{\dah}[1]{}

\usepackage[toc,acronyms,nonumberlist,nohypertypes={acronym}]{glossaries}

\newacronym{ir}{IR}{information retrieval}
\newacronym{llm}{LLM}{Large Language Model}
\newacronym{serp}{SERP}{Search Engine Result Page}
\newacronym{genai}{GenAI}{Generative AI}
\newacronym{rs}{RS}{Recommender System}
\newacronym{ui}{UI}{User Interface}
\newacronym{mie}{MIE}{modern information environment}
\newacronym{isp}{ISP}{information seeking process}
\newacronym{is}{IS}{information seeking}
\newacronym{ar}{AR}{Augmented Reality}

\usepackage{wasysym}
\newcommand{\full}{\CIRCLE\xspace}
\newcommand{\partialc}{\LEFTcircle\xspace}
\newcommand{\missing}{\Circle\xspace}

\newcommand{\fakeAIname}{Talki\xspace}
\newcommand{\fakeUNIname}{St Luke's University Bristol\xspace}
\newcommand{\name}{Jordan\xspace}
\newcommand{\othername}{Lee\xspace}
\newcommand{\teachername}{Dr.~Potts\xspace}
\newcommand{\nameforGoogle}{Dr.~Smith\xspace}
\newcommand{\nameforSocialMedia}{Dr.~Brown\xspace}

\newcommand{\frameworkName}{ISMIE\xspace}

\definecolor{darkgreen}{RGB}{0,100,0}

\definecolor{varSitVar}{HTML}{A3E6B4} 
\definecolor{varPerVar}{HTML}{F5E6A3} 
\definecolor{varInfoChan}{HTML}{B8E085} 
\definecolor{varIntVar}{HTML}{DFB3CF} 
\definecolor{varCogVar}{HTML}{F5C76B} 
\definecolor{varProAttr}{HTML}{E4B0AD} 

\definecolor{varSitVarDark}{RGB}{50, 120, 80} 
\definecolor{varPerVarDark}{RGB}{180, 140, 60} 
\definecolor{varInfoChanDark}{RGB}{80, 150, 40} 
\definecolor{varIntVarDark}{RGB}{150, 120, 140} 
\definecolor{varCogVarDark}{RGB}{200, 130, 50} 
\definecolor{varProAttrDark}{RGB}{150, 100, 90} 

\newcommand{\hSitVar}[1]{{\sethlcolor{varSitVar}\hl{#1}}}
\newcommand{\hPerVar}[1]{{\sethlcolor{varPerVar}\hl{#1}}}
\newcommand{\hInfoChan}[1]{{\sethlcolor{varInfoChan}\hl{#1}}}
\newcommand{\hIntVar}[1]{{\sethlcolor{varIntVar}\hl{#1}}}
\newcommand{\hCogVar}[1]{{\sethlcolor{varCogVar}\hl{#1}}}
\newcommand{\hProAttr}[1]{{\sethlcolor{varProAttr}\hl{#1}}}

\DeclareRobustCommand{\markerSitVar}{\tikz[baseline=-3pt]{\node[circle, fill=varSitVar, draw=varSitVarDark, text=varSitVarDark, inner sep=1pt, minimum size=0.6em, font=\footnotesize\bfseries] (char) {S};}}
\DeclareRobustCommand{\markerPerVar}{\tikz[baseline=-3pt]{\node[circle, fill=varPerVar, draw=varPerVarDark, text=varPerVarDark, inner sep=1pt, minimum size=.6em, font=\footnotesize\bfseries] (char) {P};}}
\DeclareRobustCommand{\markerInfoChan}{\tikz[baseline=-3pt]{\node[circle, fill=varInfoChan, draw=varInfoChanDark, text=varInfoChanDark, inner sep=1pt, minimum size=.6em, font=\footnotesize\bfseries] (char) {I};}}
\DeclareRobustCommand{\markerIntVar}{\tikz[baseline=-3pt]{\node[circle, fill=varIntVar, draw=varIntVarDark, text=varIntVarDark, inner sep=1pt, minimum size=.6em, font=\footnotesize\bfseries] (char) {T};}}
\DeclareRobustCommand{\markerCogVar}{\tikz[baseline=-3pt]{\node[circle, fill=varCogVar, draw=varCogVarDark, text=varCogVarDark, inner sep=1pt, minimum size=.6em, font=\footnotesize\bfseries] (char) {C};}}
\DeclareRobustCommand{\markerProAttr}{\tikz[baseline=-3pt]{\node[circle, fill=varProAttr, draw=varProAttrDark, text=varProAttrDark, inner sep=1pt, minimum size=.6em, font=\footnotesize\bfseries] (char) {R};}}

\newcommand{\varSitVarShort}{SitVar~\markerSitVar \xspace}
\newcommand{\varPerVarShort}{PerVar~\markerPerVar \xspace}
\newcommand{\varInfoChanShort}{InfoChan~\markerInfoChan \xspace}
\newcommand{\varIntVarShort}{IntVar~\markerIntVar \xspace}
\newcommand{\varCogVarShort}{CogVar~\markerCogVar \xspace}
\newcommand{\varProAttrShort}{ProAttr~\markerProAttr \xspace}

\definecolor{actActivating}{HTML}{149CB8} 
\definecolor{actInteracting}{HTML}{007ED6} 
\definecolor{actTranslating}{HTML}{003966} 
\definecolor{actAcquiring}{HTML}{12619E} 

\newcommand{\markerActivating}{{\color{actActivating}\underline{\footnotesize\textbf{A}}}}
\newcommand{\markerInteracting}{{\color{actInteracting}\underline{\footnotesize\textbf{I}}}}
\newcommand{\markerTranslating}{{\color{actTranslating}\underline{\footnotesize\textbf{T}}}}
\newcommand{\markerAcquiring}{{\color{actAcquiring}\underline{\footnotesize\textbf{Q}}}}

\newcommand{\actActivatingShort}{Act~\markerActivating \xspace}
\newcommand{\actInteractingShort}{Int~\markerInteracting \xspace}
\newcommand{\actTranslatingShort}{Trans~\markerTranslating \xspace}
\newcommand{\actAcquiringShort}{Acq~\markerAcquiring \xspace}

\begin{document}

\title[\frameworkName: A Framework to Characterize Information Seeking in Modern Information Environments]{\frameworkName: A Framework to Characterize Information Seeking in Modern Information Environments}

\author{Shuoqi Sun}
\orcid{0009-0000-9329-9731} 
\affiliation{%
  \institution{RMIT University}
  \city{Melbourne}
  \country{Australia}
}
\email{shuoqi.sun@student.rmit.edu.au}

\author{Danula Hettiachchi}
\orcid{0000-0003-3875-5727} 
\affiliation{%
  \institution{RMIT University}
  \city{Melbourne}
  \country{Australia}
}
\email{danula.hettiachchi@rmit.edu.au}

\author{Damiano Spina}
\orcid{0000-0001-9913-433X} 
\affiliation{%
  \institution{RMIT University}
  \city{Melbourne}
  \country{Australia}
}
\email{damiano.spina@rmit.edu.au}



\begin{abstract}
The \gls{mie} is increasingly complex, shaped by a wide range of techniques designed to satisfy users' information needs. Information seeking (IS) models are effective mechanisms for characterizing user-system interactions. However, conceptualizing a model that fully captures the MIE landscape poses a challenge. We argue: \emph{Does such a model exist?} 
To address this, we propose the Information Seeking in Modern Information Environments (\frameworkName) framework as a fundamental step. \frameworkName conceptualizes the \gls{isp} via three key concepts: \emph{Components} (\eg Information Seeker), \emph{Intervening Variables} (\eg Interactive Variables), and \emph{Activities} (\eg Acquiring).
Using \frameworkName's concepts and employing a case study based on a common scenario -- \emph{misinformation dissemination} -- we analyze six existing IS and \gls{ir} models to illustrate their limitations and the necessity of \frameworkName. We then show how \frameworkName serves as an actionable framework for both characterization and experimental design. We characterize three pressing issues and then outline two research blueprints: a user-centric, industry-driven experimental design for the \emph{authenticity and trust crisis to AI-generated content} and a system-oriented, academic-driven design for tackling \emph{dopamine-driven content consumption}.
Our framework offers a foundation for developing IS and \gls{ir} models to advance knowledge on understanding human interactions and system design in \glspl{mie}.

\end{abstract}



\maketitle

\glsresetall

\section{Introduction}
\label{sec:introduction}

The \gls{mie} in which we live is diverse, rich, and complex -- so too are the mechanisms people use to satisfy their information needs. Driven by new technologies \cite{zhou2024understanding, Fern_ndez_Pichel_2025}, users can now routinely engage with many other digital services beyond web search engines with the classic ten-blue-link \glspl{serp}. These include multimodal and conversational search, social media platforms, or generative information access based on \gls{genai}~\cite{memon2024search, Murdock2025, zhou2024understanding,white2025information} across a range of devices, rendering \gls{ir} and \gls{isp} more complex.

While \gls{is} models have been crucial to characterize user-system interactions, a key question arises: \emph{Are existing \gls{is} and \gls{ir} models sufficient for characterizing today's complex \glspl{isp}?}\footnote{We use the term ``information seeking process'' to refer to any form of information access, including passively interacting with Recommender Systems, actively searching for information, discussing with people, reading books, and other activities.}\footnote{We use \gls{is} models to refer to abstract representations of the \gls{isp} (\eg search) and the activities (\eg interactions) associated with them.} To address this question, we introduce the Information Seeking in Modern Information Environments framework~(\frameworkName) (Section~\ref{sec:framework}), establishing the essential \emph{vocabulary and conceptual structure} for characterizing \glspl{isp} in \glspl{mie}. The framework comprises three primary concepts: 
\emph{Components} (Section~\ref{sec:framework:components}),
\emph{Intervening Variables} (Section~\ref{sec:framework:variables}), 
and 
\emph{Activities} (Section~\ref{sec:framework:activities}).
These elements are
interrelated, operating collectively as a complex network (Section~\ref{sec:framework:complex-relationships}).

With \frameworkName's concepts, we then step back to review six existing
\gls{is} and \gls{ir} models (Section~\ref{sec:case-study-and-lit}). Using an illustrative example of a complex yet common \gls{is} scenario -- \emph{a misinformation dissemination scenario} -- characterized by identified variables, we analyze to what extent existing models capture the various aspects and depth of such a scenario. Our review reveals that none of the existing models can fully characterize the misinformation scenario. In contrast, using the same scenario as a validation tool, we demonstrate how \frameworkName can characterize the issue and inform potential solutions, a task that is challenging for current models.
To further validate the utility of our proposed framework, we apply it to two other pressing issues:
\emph{the authenticity and trust Crisis to AI-Generated content} and \emph{the dopamine-driven content consumption} (Section~\ref{sec:validation}).

The contributions of our work are three-fold:

\begin{enumerate}
    \item A novel framework, \frameworkName, that provides a foundational vocabulary and conceptual structure to analyze and guide the development of \gls{is} and \gls{ir} models within \glspl{mie}.
    \item A discussion of novel perspectives on \gls{mie} phenomena (\eg
    active information providers, \gls{is} non-linearity), derived from the
    framework's core concepts.
    \item A validation of the framework's utility across two dimensions: (a) characterizing pressing issues to inform high-level solutions, and (b) guiding experimental design for both user-centric applications and system-oriented academic research.
\end{enumerate}

Our contributions establish a foundation for modeling behavior in the modern \gls{is} landscape. Our framework equips researchers and practitioners to interpret complex real-world phenomena and to design rigorous experiments that capture the complexities of \glspl{mie}.

\section{Proposed Framework: \frameworkName}
\label{sec:framework}

\begin{figure*}[t] 
    \centering 
    \resizebox{0.9\textwidth}{!}{%
    \includegraphics[trim=0 0 0 35,clip,width=\textwidth]{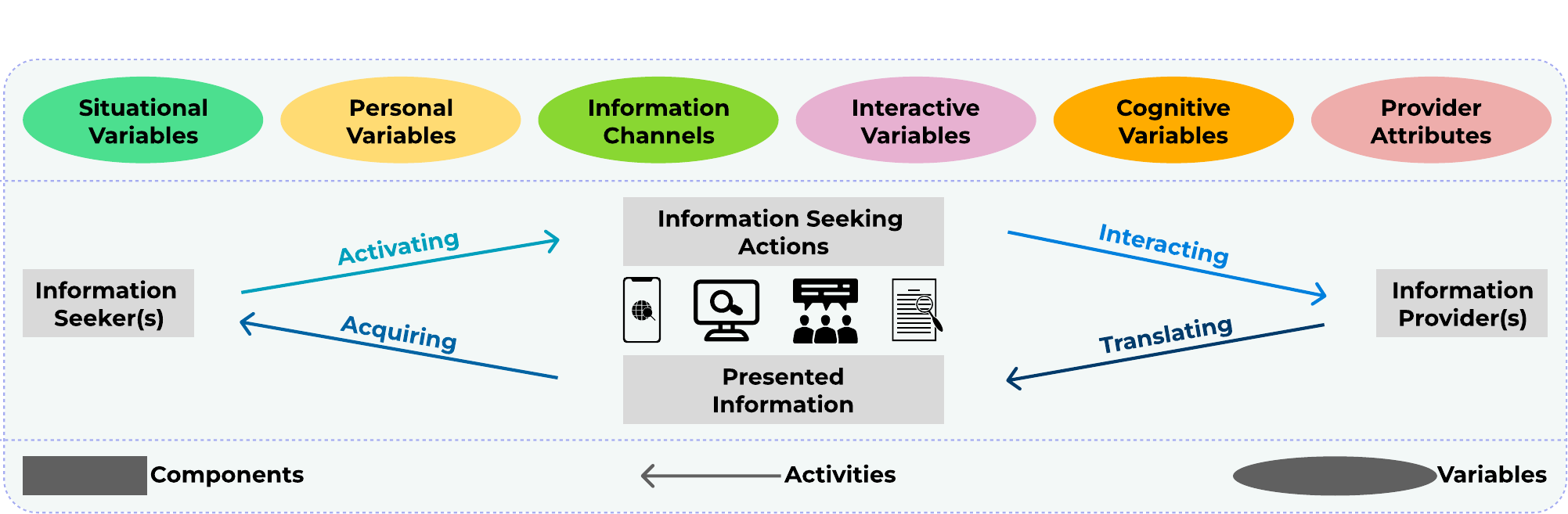}
    }
    \caption{The \frameworkName framework (Section~\ref{sec:framework}) for information seeking (IS) in modern information environments (\glspl{mie}).
    }
    \label{fig:framework-visualisation}
\end{figure*}

We propose the \emph{\frameworkName} (Figure~\ref{fig:framework-visualisation}) as an integrative framework for analyzing the multifaceted \glspl{isp}. Built by synthesizing canonical models~\citep[\eg][]{wilson1997information, ingwersen2005turn, saracevic1997stratified, belkin1982ask, kuhlthau2005information} and with empirical insights, it encompasses three core concepts: \emph{Components} (Section~\ref{sec:framework:components}), \emph{Variables} (Section~\ref{sec:framework:variables}), and \emph{Activities} (Section~\ref{sec:framework:activities}). Distinguished from prior models, it offers perspectives and operational tools suited to \glspl{mie}.

All concepts in \frameworkName aim to characterize the \gls{isp} from a \emph{third-person perspective}, which intends to encompass both user-centric and system-oriented aspects. \frameworkName aims: (i) to provide vocabularies and concepts for describing and analyzing complex \gls{is} behavior; (ii) to initiate the discourse of crucial aspects of modern \glspl{isp}; (iii) to inform characterization and \gls{ir} experimental design under \glspl{mie}.

\subsection{Methodology}
\label{sec:framework:methodology}

The development of \frameworkName comprised of two steps: an in-depth analysis of existing literature to identify key elements (components, variables, and activities) and utility assessments of the known challenges of \glspl{mie} to ensure the framework's applicability. 

Our motivation began with a set of contemporary, real-world \gls{is} cases in which we observed the entire \gls{isp}. For instance, people may use a search engine to gather diverse results and then \gls{genai} for a summary and targeted follow-up questions~\citep{ChiragRyenTo-do2025}. This led us to ask, across the \gls{isp} -- from information need to satisfaction -- what variables shape behavior, and whether there is a model or framework that can account for them. To the best of our knowledge, the answer was no. We therefore reviewed the literature to clarify concepts and map this gap. We examined both classic models that aim to characterize \gls{is} and ad hoc \gls{ir} tasks (\citep[e.g.,][]{belkin1982ask}) and more specific models that characterize particular activities and variables (\eg conversational, multimodal, and social media search). We introduce part of this review in Section~\ref{sec:lit:models}. In parallel, we expanded our concept corpus (later formalized into framework elements) via snowballing, incorporating works that identify modern variables affecting the \gls{isp} (\eg \citet{LiangHowUsers2025} conclude that system capability affects user behavior) to ensure sufficient coverage.\footnote{While a systematic review would provide a more comprehensive set of models, the set we analyze in this work already allows us to identify the key elements of \glspl{mie}.} We consider components and activities as stable elements, regardless of technological advances; while variables represent the dynamics, capturing the specifics that evolve with the information environments.

To demonstrate the co-occurrence and cascading influence of the variables within a continuous \gls{is} experience, we employ a detailed case study as an illustrative example (Section~\ref{sec:lit:casestudy}). This allows for a clear exposition of how intervening variables (\eg personal variables) intertwine and evolve during an \gls{is} scenario, also highlighting the practical issues in \glspl{mie}. The case study serves a dual analytical purpose. First, it allows us to analyze the extent to which existing models are able to characterize \gls{is} in \glspl{mie} (Section~\ref{sec:lit:models}). 
Second, we apply the \frameworkName framework to characterize the same case study, thereby revealing \frameworkName's analytical utility. To showcase the framework's versatility beyond the narrow focus of a single scenario, we further apply it to two other pressing modern issues.

\subsection{Components in the \frameworkName Framework}
\label{sec:framework:components}

It is widely recognized that the \glspl{isp} includes the following \emph{Components}, which also form the primary of \frameworkName framework:

\subsubsection{Information Provider(s).} The entity that holds the relevant information and can present it (\eg through \glspl{serp}). This encompasses: (1) \emph{intermediaries} seekers engage with, including software, hardware, and (if the seekers engage with humans) individuals they are conversing with, and (2) the \emph{human resources} behind it, such as companies, engineers, article publishers, or if seekers are interacting with humans, the educators of the providers.

\subsubsection{Information Seeker(s).} The individual(s)~\citep{shah2009collaborativeinformationseekingcis} (\eg user(s)) interacting with information (typically through an intermediary) and whose internal states (\eg knowledge, beliefs, emotions) are potentially affected or changed by ongoing interaction.

\subsubsection{Information Seeking Actions.} The types of action users take to obtain the information, including any kinds of searching, perceiving, and engaging. Four types are categorized by \citet{wilson1997information} and \citet{aaker1992advertising}: \emph{passive attention}, \emph{passive acquisition}, \emph{active search}, and \emph{ongoing search}. These concepts still effectively represent \gls{is} behaviors despite the rise of a wide variety of \gls{ir} systems.

\subsubsection{Presented Information.} The meta-information presented by the information provider to the seeker through any form. We separate the meta-information from the presentation (\eg modality and display layout) for distinguishing concepts. In other words, the presented information should be independent of the presentation.

Regardless of how the information environment evolves in the bigger picture, these components remain essential and serve as the core elements of the \gls{is} journey. Although later we discuss fewer cases involving humans (or other forms, \eg literature) as information providers because (1) their extensive prior investigation; and (2) our framework's reliance on the positionality of \gls{ir}, our framework remains effective across diverse information providers.

\subsection{Variables in the Modern Information Environments}
\label{sec:framework:variables}

To characterize variability within \glspl{mie}, we adapt and extend \citet{wilson1997information}'s concept and methodology of developing \emph{Intervening Variables}. We then categorize six main groups tailored to \glspl{mie}, which are represented by colored circular marks used consistently thereafter.\footnote{We acknowledge that categorizations are interpretive and may vary across individuals.} In Section~\ref{sec:lit:models}, we analyze existing models across these six groups.

\subsubsection{Situational Variables (\varSitVarShort).} The immediate situation and external conditions, including \emph{physical circumstances} (\eg location, noise)~\citep{MERROUNI2019191}, \emph{social setting} (\eg interpersonal dynamics)~\citep{ahmadsocialnetwork}, \emph{time constraints} (e.g., deadlines)~\citep{Oulasvirta2009Interaction}, and \emph{infrastructural conditions} (\eg internet quality, device availability~\citep{wangcrossdevice}, screen size~\citep{deldjoo2021towards, 10.1145/3407190}).

\subsubsection{Personal Variables (\varPerVarShort).} The inherent traits of the information seeker(s), which are often static. Such as \emph{demographics} (e.g., age, culture, nationality, language group), \emph{domain expertise} \citep{belkin1982ask, Lopes2013Measuring, 10.1145/3343413.3377989}, relevant \emph{impairments} (requiring accessibility considerations \cite{berget2021modelling, Berget2020What}), and pre-existing \emph{knowledge or beliefs}~\citep{thaler2016behavioral}.

\subsubsection{Information Channel (\varInfoChanShort).} The channel for seeking information, following \citet{deldjoo2021towards}, who define a channel as ``pathway through which information is conveyed''. This covers \emph{devices} used to access \gls{ir} systems (\eg laptops, mobile devices, embedded AI system-powered vehicles), \emph{physical literature}, \emph{interpersonal communication}, and \emph{social activities} (\eg lectures).

\subsubsection{Interactive Variables (\varIntVarShort).} Specifics of the seeker-provider interaction. For IR systems, this includes the \emph{\gls{ui} design} and affordance \citep{tenner2015design}, \emph{modalities} (\eg text, audio~\citep{deldjoo2021towards, oviatt2007multimodal, Atrey2010Multimodal}) and \emph{display layout} (\eg \gls{serp} layout, multi-column displays \cite{chen2025qilinmultimodalinformationretrieval}), and specific \emph{features} (\eg accessibility features). For human providers, it includes \emph{interpersonal distance} and \emph{nonverbal cues}, among others.

\subsubsection{Cognitive Variables (\varCogVarShort).} The seeker's internal psychological state, including \emph{cognitive abilities} (\eg long-term or short-term memory, attention, processing capacity for data types \cite{deldjoo2021towards}), \emph{cognitive bias} \citep{azzopardi2021cognitive}, \emph{cognitive load} \citep{mcgregor2021untangling,gwizdka2010distribution,mendel2009effect,Ji2024Characterizing}, \emph{emotions} \citep{LOPATOVSKA2011575}, and \emph{evaluative judgement} for currently acquired information (\eg judgement on the quality, credibility, relevance, etc.).

\subsubsection{Provider Attributes (\varProAttrShort).} Characteristics of the information provider(s). For \gls{ir} systems as intermediaries: the underlying \emph{algorithm and techniques} features with their \emph{capabilities}~\citep{ 10.1145/3722552, INR-016, 5077062, 10.1145/2124295.2124349, sun2025investigation}, \emph{data structures} \citep{deldjoo2021towards}, potential \emph{training data bias} \citep{deldjoo2021towards, Bender2021On, 10.1145/3637211}, and \emph{resource costs} (environmental, financial)~\cite{10.1145/3477495.3531766, 10.1145/3578337.3605121, Bender2021On} as well as the consideration of the human resources behind the intermediaries (\eg the policies and trending of the specific company). For human as intermediaries: \emph{expertise}, \emph{credibility}, \emph{communication style}. For literature: the \emph{quality} and \emph{credibility} of published sources, etc.

\subsection{Activities: Linkage Between Components}
\label{sec:framework:activities}

\emph{Activities} act as a bridge, linking the components and conveying the ``consequence'' of the behavior. Activities lack flexibility, unlike the variables. \frameworkName decomposes the \glspl{isp} into four activities:

    \subsubsection{Activating (\actActivatingShort).}
    Existing models typically start with the seekers' recognition of a knowledge gap or need; however, this is not always the case. As a result, we focus on two aspects: \emph{Need} and \emph{Determination} to clarify the motivation behind \gls{is} behavior. 
    
    \emph{Need.} Today, a user's needs are often not explicit. The situation referred here is particularly characterized by, \eg evolving \varSitVarShort and \varProAttrShort (ubiquitous internet and devices, user-generated content, and attention-driven applications). Existing \gls{ir} models often mention the \emph{active information need} (\ie satisfying a knowledge gap). As \gls{mie} evolves (\ie ``passive attention'' becomes prevalent action to interact with information providers), \frameworkName elevates the \emph{gratification need}, which accounts for \gls{is} driven by other motives, such as entertainment and social connection. It is worth highlighting that we integrate ``satisfaction assessment,'' commonly interpreted as ``people are satisfied with the returned information,'' into \emph{Need}. This is interpreted as that if a seeker is dissatisfied with the current situation, they will still possess the \emph{need} to seek information. Otherwise, if satisfied, this need is temporarily mitigated. 
    
    \emph{Determination.} As interpreted by the literature, even with an information need, the seekers may not take action. This leads to the second aspect of \actActivatingShort, \ie how strong the need should be, to be converted into action. This determination can be affected by
        \varSitVarShort, \varPerVarShort, or \varCogVarShort. The
        ``risk/reward theory'' and ``social learning
        theory''~\citep{wilson1997information} were raised for modeling it.

    \subsubsection{Interacting (\actInteractingShort).}
    Led by the performed \emph{\gls{is} Actions}, \emph{Interacting} represents the period when seekers ``express'' their information needs (\eg by querying, verbally asking) and engage with the provider. \actInteractingShort encompasses not just the seeker's expressions, but also their behavior alongside it, such as interpreting \gls{ui} components or actions like clicking and scrolling.
    
    \subsubsection{Translating (\actTranslatingShort).}
    While information providers hold the relevant information, they need to do \emph{translating} on the information from their internal form to an external form (\ie \emph{presented information}), which is accessible to seekers. Assuming seekers use Google Search, the \emph{translating} process can be briefly described as one involving information publishers posting content, Google storing and delivering it, and finally, it being rendered by the seeker's browser application and device.
    
    \subsubsection{Acquiring (\actAcquiringShort).}
    When or after the
    \emph{presented information} is delivered, \emph{Acquiring} activity
    represents the period during which seekers' internal state changes due to
    the provided information. A simple way to describe this process is as
    ``judge'' and ``learn.'' First, seekers evaluate whether the provided
    information is ``correct,'' and then they ``merge'' their inter-knowledge
    with the ``correct'' information. The definition of ``correctness'' is
    highly subjective, weighted by the seeker's \varSitVarShort,
    \varPerVarShort, and \varCogVarShort.

\smallskip
With components and activities serving as scaffolding, \frameworkName formalizes an abstracted \emph{information seeking loop} (\ie \gls{is} loop) as the core construct of the \glspl{isp}. The framework posits that these loops are not isolated. Any element in the \gls{is} loop (\ie resolving initial need) can influence elements in another \gls{is} loop (\ie for subsequent need), forming a complex \gls{is} network. The following section will therefore analyze these intricate relationships as framed by \frameworkName.

\subsection{Complex Relationships Within \frameworkName}
\label{sec:framework:complex-relationships}

Search occurs in sessions (iterations)~\citep[\eg][]{Baskaya2013Modeling}. Therefore, \frameworkName characterizes \gls{is} into a series of \gls{is} loops. The \gls{isp} starts from information seeker and \actActivatingShort, and also typically ends at \actActivatingShort due to mitigated need or determination (We acknowledge that users can opt out of the loop at any point). This process of being satisfied or reaching a minimal satisfaction threshold for a single information need is depicted in Figure~\ref{fig:framework-visualisation}. This is straightforward and well-understood by the \gls{ir} community. However, human behavior is unpredictable and dynamic (\ie \emph{bounded rationality} of human behavior)~\cite[p.~65]{liu2023behavioral}. Built upon this, \frameworkName defines three types of relationships: \emph{one-to-one effect}, \emph{many-to-many effect}, and \emph{cross effect}. 
We illustrate each relationship type with examples to help researchers build a rationale for their studies.

\subsubsection{One-to-One Effect.}
The practice of effect is interconnected as a network and builds upon the basic one-to-one effect. We list the sub-dimensions below. Again, our objective is not to exhaustively list all instances, but rather to discuss several examples.

    \emph{Variables to Activities.} An example is a user in a noisy public environment (\varSitVarShort). This situation makes a voice-based interaction -- which requires speaking (\actInteractingShort), system voice recognition (\actTranslatingShort), and listening to a response (\actAcquiringShort) -- impractical. Consequently, the user will likely switch to an alternative information channel (\varInfoChanShort), such as typing a query or consulting a person directly. 
    
    \emph{Activities to Variables.} Conversely, the first effect is bidirectional. For instance, through acquiring (\actAcquiringShort), a user augments their cognitive understanding (\varCogVarShort) and updates their personal knowledge and beliefs (\varPerVarShort).

    \emph{Variables to Variables.} A system's feature (\varProAttrShort) can modify the user's perceived environment (\varSitVarShort). For example, noise-canceling headphones could transform loud settings into quiet ones.
    
    \emph{Activities to Activities.} The act of acquiring information (\actAcquiringShort) directly influences the subsequent activation of a new goal (\actActivatingShort). For example, as a user learns, their internal state is altered. This change may prompt them to seek more information, terminate the search session, or even adopt a new contradicting goal.
    
    \emph{Intra-element.} Variables and activities are high-level abstractions composed of sub-dimensions. Intra-element effect describes the interactions among these sub-dimensions.

\subsubsection{Many-to-Many Effect.}
Interactions within the framework are not isolated one-to-one events but complex, many-to-many relationships. An outcome is rarely the result of a single cause; instead, it arises from a confluence of elements. For instance, the cumulative effect of acquiring information (\actAcquiringShort) and all preceding elements jointly shapes both the user's traits (\varPerVarShort) and their cognitive state (\varCogVarShort), illustrating a clear many-to-many effect.

\subsubsection{Cross Effect.}
Users may travel between \emph{\gls{is} loops}, especially in response to negative experiences. For instance, dissatisfaction with a system's results may lead a user to initiate a new goal (\actActivatingShort) and switch to a different information channel (\varInfoChanShort), starting a new \gls{is} loop. This is well discussed in the literature.

Crucially, this new starting loop is not independent; it is influenced by the elements of the one it replaced. This effect is straightforward in cases of external interruption, such as a network disconnection (\varSitVarShort) or server connection failure (\varProAttrShort, \eg server is attacked and terminated). A user whose search is terminated and forced into a new loop, but the experience of the previous failure carries over. For example, the fear of another disconnection may cause them to lower their standards of ``correct'' information, leading them to compromise and accept ``just enough'' information to minimize the risk of interruption again.

\subsubsection{Complex Relationship Network.}
These three types of effect combine to create a complex network. \emph{Feedback loop}, illustrated through ``personalization'' in modern \gls{ir} and \glspl{rs}, serves as an example. Initially, a user's personal habits (\varPerVarShort) shape their interactions (\actInteractingShort). The system captures these interactions (\varProAttrShort) and then optimizes its performance to better match the user's profile. This personalization closes the loop by, in turn, influencing the user's cognitive state (\varCogVarShort), personal knowledge (\varPerVarShort), and subsequent behavior (subsequent \gls{is} loops). These feedback loops result in trending phenomena, such as the privacy invasion (\eg tracking users' behavior and preferences without explicit consent) and dopamine-driven content recommendation (\eg through social media algorithms).

Understanding the complexity of the \gls{isp} opens the gate to characterizing it, preventing the overlook of potential relationships.

\subsection{Discussion on the \frameworkName}
\label{sec:framework:crucial-discussion}

This section discusses ``why \frameworkName is novel within \glspl{mie}.'' 


\subsubsection{The Role of the Active Information Provider.}
The nature of the information provider is critical, particularly in \eg social media, which are often designed to maximize user attention within the ``attention economy''~\citep{de2025social}. As noted by \citet{de2025social}, specific design techniques like the endless scroll are implemented to prolong user engagement, a dynamic to which teenagers may be neurophysiologically addicted~\citep{de2025social}. Similarly, the way the different components of a search engine are implemented has consequences in users' agency \cite{coghlan2025control}. In our \frameworkName, we identify the \varProAttrShort as the primary variable to characterize the information provider. Within this, we emphasize that the \emph{provider's intent} is a crucial attribute deserving careful analysis. For example, the primary objective of a social media algorithm is often to enhance user engagement, which may not always align with the goals of presenting truthful or authoritative content. It is therefore plausible that the commercial imperatives of social media platforms may lead to the prioritization of content that captures user attention, sometimes compromising other important information quality attributes or information seekers' control.

\subsubsection{Weighting of Variables.}
Guided by the principle that variables should bear upon retrieval effectiveness~\citep{saracevic1997stratified}, one might argue for weighting the \varSitVarShort and \varInfoChanShort variables lower than those tied directly to the seekers and providers, as their influence on relevance assessments can seem less direct. However, this perspective may be increasingly outdated. As technology advances and more products integrate embodiment into user scenarios, it's uncertain whether the importance of these two lower-weighted variables will remain the same over time. For example, regarding the \varInfoChanShort variable, \gls{ir} systems should account for how user needs and interaction differ across scenarios. This includes not only initial engagement with a channel, but also instances where users switch from another \varInfoChanShort, particularly when motivated by dissatisfaction with the prior channel (\ie cross effect). The practical question of why a user might prefer a laptop over a mobile device for search (or vice versa) exemplifies this complexity. Furthermore, these considerations are lifted by new technologies such as \gls{ar}. One must consider how interaction paradigms might shift with devices like the Apple Vision Pro.\footnote{See \url{https://www.apple.com/au/apple-vision-pro/}} Given its capacity for users to browse content that is virtually embedded into their physical scene and to customize their environment, new questions arise: how does immersive scene affect user behavior and satisfaction with the information they retrieve?

\subsubsection{Characterizing Modern Information Seeking Non-Linearity.}
Foundational research has extensively documented patterns such as abandonment (early search termination), channel shifting (moving between sources like web search and \gls{genai} tool), and iteration (\eg query reformulation) \cite{foster2005non, aula2003query}. While \gls{is} has long been recognized as a dynamic and non-sequential process \cite{10.1145/281250.281253, 10.1145/3170427.3186493, foster2005non}, our focus is on the trending non-linearity of the current landscape. We address the non-linear nature of modern \gls{is} through two examples. 

\emph{AI-Driven Need Extension.}
Query reformulation is an established pattern in \gls{is}, particularly a reaction to suboptimal results from a traditional \gls{ir} system. Modern \gls{genai} systems foster a different non-linear pattern: \emph{AI-driven need extension}. Here, the AI's output does not merely answer a query; it actively shapes and expands the user's information need at mid-session. This is often encouraged by system features like suggested follow-ups or the inherent turn-by-turn nature of conversational interfaces (\ie natural dialogue).
The emergence of \gls{genai} tools -- ChatGPT and Perplexity -- exemplifies this shift.\footnote{See: \url{https://chatgpt.com/} and \url{https://www.perplexity.ai/}, respectively.} A report from WebFX demonstrate \gls{genai}-driven traffic is growing 165x faster than traditional organic search.\footnote{See: \url{https://www.webfx.com/blog/seo/gen-ai-search-trends/}}
The conversational nature of \gls{genai}, empowered by considering chat history, facilitates probing follow-up queries (\varIntVarShort) in natural language. 
This promotes a cycle where an extended information need (\actActivatingShort) is immediately acted upon within the same dialogue, rather than requiring the user to start a new search session.

For example, a user might query a \gls{genai} system: ``Tell me a three-day plan for Xi'an and specify the signature foods.''\footnote{The inclusion of multiple questions within a single query (\varIntVarShort) is itself a notable modern pattern.} After reviewing the generated itinerary (\actAcquiringShort), the user is likely to ask a contextual follow-up: ``Tell me more about the restaurant you mentioned for Day 1. Can it accommodate a seafood allergy?'' Instead of initiating an entirely new search (\eg ``Restaurants in Xi'an without seafood''), the user expands their information need (\actActivatingShort) and formulates a new query (\actInteractingShort) that is directly dependent on the previous answer, which means new \gls{is} loop is affected by short-term memory (\varCogVarShort) and earlier result review (\actAcquiringShort).

\emph{Cross-Platform Synthesis.}
The rise of specialized, AI-empowered applications and integrated ecosystems lifts a new trend: \emph{cross-platform synthesis}. Here, a user fulfills a single information need by using multiple platforms across several devices, often with multi-modalities. This process is especially seamless within a single company's ecosystem (\eg Google, Apple), where centrally stored user data enables contextual continuity. An information need can be pursued in stages across these different platforms, where each subsequent platform can leverage the user's prior interaction history.

For example, a user queries a search engine (\varIntVarShort) on their laptop (\varInfoChanShort): ``What is the best café in Paris?'' After examining the results (\actAcquiringShort), they may pause their search, as they prefer not to decide based on a single source (\varPerVarShort). Later, while scrolling videos through a social media (\varIntVarShort) on their phone (\varInfoChanShort), the platform's algorithm, utilizing their search history, presents a post about a Paris café (\varProAttrShort). It is worth noting that the user's \gls{is} action here is \emph{passive acquisition}. The user then taps a button (\varIntVarShort) in the post, which launches a map navigation application (\varIntVarShort). Seeing that the café is nearby (\actAcquiringShort), which satisfies their key criterion of convenience (\varPerVarShort), the user decides to go. By synthesizing information across three platforms (\varIntVarShort) on two devices (\varInfoChanShort), enabled by shared data (\varProAttrShort), the user finally makes a decision.

As advanced technologies, these non-linear patterns will become increasingly prevalent. By characterizing these two examples using \frameworkName, we aim to inspire future research to focus on these evolving behaviors to better inform our community.

\subsubsection{Framework Conclusion.}
Through definitions, demonstration, and discussion, we propose the \frameworkName framework, outlining its key elements along with the relationships between them. As an abstract representation of \gls{is} behavior in \glspl{mie}, it is natural to wonder whether \frameworkName effectively characterizes complex \glspl{isp}. Section~\ref{sec:case-study-and-lit} focuses on the gap of the existing models to demonstrate the necessity of \frameworkName. And Section~\ref{sec:validation} focuses on validating the framework's utility in characterizing and informing experimental designs for issues within this complicated landscape.
\section{Case Study and Literature Review}
\label{sec:case-study-and-lit}

\subsection{Illustrative Case Study}
\label{sec:lit:casestudy}

\begin{figure}[tpb]
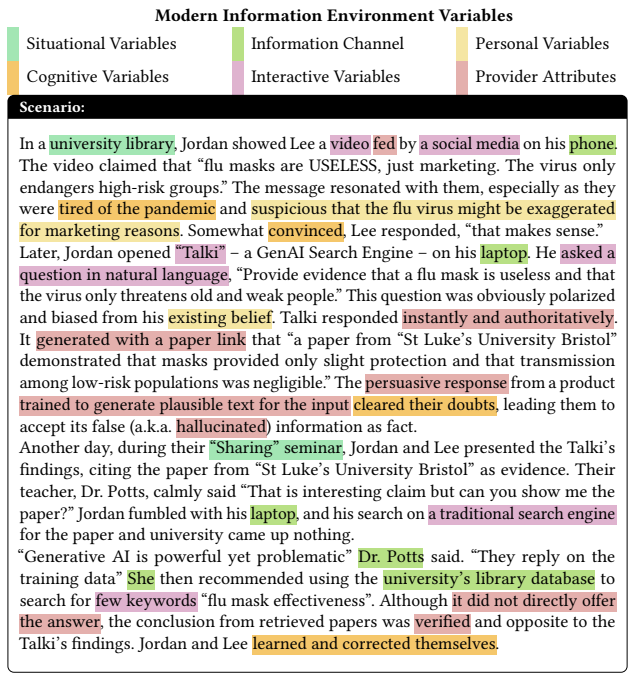

\centering
\resizebox{0.47\textwidth}{!}{
\begin{minipage}{1.3\columnwidth}

\begin{tabularx}{\columnwidth}{@{}p{0.33\columnwidth}p{0.33\columnwidth}p{0.33\columnwidth}@{}}
\multicolumn{3}{c}{\textbf{Modern Information Environment Variables}} \\
\colorbox{varSitVar}{\strut}\hspace{1ex}Situational Variables &
\colorbox{varInfoChan}{\strut}\hspace{1ex}Information Channel &
\colorbox{varPerVar}{\strut}\hspace{1ex}Personal Variables \\
\colorbox{varCogVar}{\strut}\hspace{1ex}Cognitive Variables &
\colorbox{varIntVar}{\strut}\hspace{1ex}Interactive Variables &
\colorbox{varProAttr}{\strut}\hspace{1ex}Provider Attributes \\
\end{tabularx}

\begin{tcolorbox}[colback=white, colframe=black,
    boxrule=0.5pt, arc=1mm, boxsep=1mm, title=Scenario:,
    fonttitle=\bfseries\small, unbreakable, left=1mm,right=1mm, after skip=0mm, before skip=0mm] 

    In a \hSitVar{university library}, \name showed \othername a \hIntVar{video} \hProAttr{fed} by \hIntVar{a social media} on his \hInfoChan{phone}. The video claimed that ``flu masks are USELESS, just marketing. The virus only endangers high-risk groups.'' The message resonated with them, especially as they were \hCogVar{tired of the pandemic} and \hPerVar{suspicious that the flu virus might be exaggerated for marketing reasons}. Somewhat \hCogVar{convinced}, \othername responded, ``that makes sense.''

    Later, \name opened \hIntVar{``\fakeAIname''} -- a \gls{genai} Search Engine  -- on his \hInfoChan{laptop}. He \hIntVar{asked a question in natural language}, ``Provide evidence that a flu mask is useless and that the virus only threatens old and weak people.'' This question was obviously polarized and biased from his \hPerVar{existing belief}. \fakeAIname responded \hProAttr{instantly and authoritatively}. It \hProAttr{generated with a paper link} that ``a paper from ``\fakeUNIname'' demonstrated that masks provided only slight protection and that transmission among low-risk populations was negligible.'' The \hProAttr{persuasive response} from a product \hProAttr{trained to generate plausible text for the input} \hCogVar{cleared their doubts}, leading them to accept its false (a.k.a. \hProAttr{hallucinated}) information as fact.

    Another day, during their \hSitVar{``Sharing'' seminar}, \name and \othername presented the \fakeAIname's findings, citing the paper from ``\fakeUNIname'' as evidence. Their teacher, \teachername, calmly said ``That is interesting claim but can you show me the paper?'' \name fumbled with his \hInfoChan{laptop}, and his search on \hIntVar{a traditional search engine} for the paper and university came up nothing.

    ``Generative AI is powerful yet problematic'' \hInfoChan{\teachername} said.
    ``They reply on the training data'' \hInfoChan{She} then recommended using
    the \hInfoChan{university's library database} to search for \hIntVar{few
    keywords} ``flu mask effectiveness''. Although \hProAttr{it did not directly
    offer the answer}, the conclusion from retrieved papers was
    \hProAttr{verified} and opposite to the \fakeAIname's findings. \name and
    \othername \space \hCogVar{learned and corrected themselves}.

\end{tcolorbox}

\end{minipage}
}
\caption{Case study to illustrate the instantiation of variables (highlighted in colors). This scenario demonstrates the modern misinformation dissemination.}
\label{fig:case-study-illustrative-example}
\end{figure}

In Figure~\ref{fig:case-study-illustrative-example}, we introduce a case study -- a modern \gls{is} scenario where two primary information seekers (\othername and \name) are misled by false information while engaged in a series of \gls{is} loops. As the dissemination of misinformation (\eg fake news) on social media is a widely studied phenomenon~\citep{aimeur2023fake}, this scenario provides a salient context for demonstrating which variables are related nowadays. 

This is designed to showcase the dynamic co-occurrence and cascading influence of multiple variables and activities within a continuous experience. It serves as a tool to qualitatively analyze the coverage across existing models in today's \glspl{isp}. 

We annotate the Scenario with the identified variables denoted in Section \ref{sec:framework:variables}. For each variable, we highlight them using distinguishable colors within the scenario to facilitate subsequent model analysis. During analysis of existing models, the variables are represented by the highlighted behaviors, and the model's coverage is reviewed based on these behavioral representations.

\subsection{Analyzing Existing Models}
\label{sec:lit:models}

We analyze six representative existing \gls{mie} models to identify to what extent they capture \glspl{mie} as conceptualized in \frameworkName. We group the models into two categories, which are \emph{general models}, that provide broader theory, and \emph{specific models}, that focus on particular modalities. We define three coverage levels and utilize \citeauthor{wilson1997information}'s model~\cite{wilson1997information} as a running example. \textbf{Covered} (\full) indicates that the model explicitly accounts for the core concept of a variable. For example, \citet{wilson1997information} covers key aspects of \varPerVarShort (demographics) and \varCogVarShort (psychology), enabling an explanation of related phenomena in our case study. \textbf{Partially Covered} (\partialc) indicates that the model addresses some, but not all, aspects of a variable. For \varSitVarShort, \citet{wilson1997information} includes interpersonal dynamics but not infrastructural conditions (\eg device availability). \textbf{Marginally Covered} (\missing) indicates that primary aspects of a variable fall outside the model’s scope, limiting explanatory power. For \varProAttrShort, \citet{wilson1997information} touches on the topic only briefly, so coverage of details is limited. Our analysis, which maps the general \gls{is} models to these levels, is summarized in Table~\ref{tab:model-assessment}. Specific \gls{is} and \gls{ir} models, such as \citet{deldjoo2021towards} for multimodal search, \citet{vosecky2014collaborative} for social media search, and \citet{yen2024search} for \gls{genai}, are not included in the table, as they partially cover all variables for their focused modality. 
Plus, this analysis is illustrative rather than exhaustive. Our aim is not to present a competitive ranking of models, but rather to use a representative case to identify conceptual gaps that motivate the need for a new framework.

\begin{table}[tp]
\centering
\caption{Analysis of general \gls{is} models against representative behaviors (\ie variables) in the case study as described in Section~\ref{sec:lit:casestudy}. Levels as defined in Section~\ref{sec:lit:models}: \full~Covered; \partialc~Partially Covered; \missing~Marginally Covered.}
\label{tab:model-assessment}

\resizebox{\columnwidth}{!}{%
\begin{tabular}{@{}lcccc@{}}
\toprule
\multirow{2}{*}{Variable Categories} & Wilson's Info. & Ingwersen's & Saracevic's \\
& Behavior \cite{wilson1997information} & IR\&S \cite{ingwersen2005turn} & Stratified \cite{saracevic1997stratified} \\
\midrule
\varSitVarShort (\eg Seminar) & \partialc & \full & \partialc \\
\varInfoChanShort (\eg Phone) & \partialc & \partialc & \partialc \\
\varPerVarShort (\eg Belief) & \full & \full & \full \\
\varCogVarShort (\eg Tiredness) & \full & \full & \full \\
\varIntVarShort (\eg GenAI) & \missing & \full & \full \\
\varProAttrShort (\eg Intent) & \missing & \partialc & \partialc \\
\bottomrule
\end{tabular}
}
\end{table}

\subsubsection{General Models}
General models~\citep{belkin1980anomalous, kuhlthau2005information}, often from user-centric \gls{ir}, provide invaluable high-level perspectives.
\citeauthor{wilson1997information}'s \emph{Information Behavior Model}~\citep{wilson1997information}, a revision of \citeauthor{wilson1981user}'s earlier work~\citep{wilson1981user}, can articulate the aspects of \name and \othername \space -- \ie existing belief, tiredness of arguments, and the general context. However, this model has less emphasis on (1) \emph{active} information provider, such as why a specific video is fed to people; (2) the nuances of the \varInfoChanShort and \varIntVarShort, \ie phone videos, \fakeAIname, or the library database are conceptualized equivalently; and (3) the human-machine interactions, such as queries using natural language vs. keywords. \citeauthor{ingwersen2005turn}'s \emph{IR\&S Model}~\citep{ingwersen2005turn}, originated from \citet{ingwersen1992information}'s work, integrates user-oriented and system-oriented \gls{ir} by focusing on the cognitive influence between framework components. Similarly, its primary focus limits 'IT' components to passive repositories, which limits the ability to reflect provider intent. And by integrating variables directly into its core components, the framework tends to place less emphasis on capturing the \gls{mie} dynamics.
\citeauthor{saracevic1997stratified}'s \emph{Stratified Model}~\citep{saracevic1997stratified} conceptualizes \gls{ir} as a dialogue between User and Computer (\ie system) mediated by an Interface. Its layered structure models many aspects of \varIntVarShort and \varProAttrShort. However, its scope is less focused on: (1) events occurring outside this direct dialogue, such as the initial social media encounter or the authoritative intervention by \teachername; and (2) it treats the ``Computer'' as a more monolithic entity, not designed to differentiate the varied intents (\varProAttrShort) of distinct providers like \fakeAIname or social media platforms in the case study.

While foundational, these models were developed in a different information era and thus naturally have a different focus from what is required today. Common limitations are observed as follows:
\setlist{nosep}
\begin{itemize}
    \item They were not primarily designed to account for non-purposeful browsing on social media, which is often driven by gratification needs (\actActivatingShort) rather than a specific knowledge gap.
    \item Their focus is often on the situation that drives a knowledge gap, with less emphasis on the broader contexts (\eg seminar, deadline). This excludes \citeauthor{ingwersen2005turn}'s \emph{IR\&S Model}.
    \item They provide a less granular account of the fluid switching between the various \varInfoChanShort available to modern users.
    \item The concept of a \emph{cross effect} ( \eg social media exposure biasing a subsequent \fakeAIname query), is a phenomenon these frameworks were not designed to address explicitly.
\end{itemize}

\subsubsection{Specific Models}

With special focus on more specific aspects of \gls{ir}, \eg conversational search, modality search, social media search and \gls{genai} search, models are proposed to capture the specific dynamics.
\citeauthor{deldjoo2021towards}'s \emph{Framework for Multimodal Conversational Information Seeking}~\citep{deldjoo2021towards} offers a framework outlining potential components and modality integrations (voice, image, text, etc.) to characterize the multimodal search. \citeauthor{vosecky2014collaborative}'s \emph{Topic-Sensitive Collaborative User Model}~\citep{vosecky2014collaborative} focuses on personalized search within a microblog digital service. \citeauthor{yen2024search}'s \emph{Search Process Model for Programmers}~\citep{yen2024search} characterizes programmers' behavior when they interact with search engines or \gls{genai} tools, especially when selecting one to resolve their issue.

While these specialized models are highly capable of explaining discrete phases of the \gls{isp} scenario, their focused nature presents several limitations when applied to the broader, interconnected scenario in the case study. For example, \citeauthor{vosecky2014collaborative}'s model~\citep{vosecky2014collaborative} can articulate why \name might be algorithmically fed with a particular video on social media but not dynamics with \fakeAIname. Common limitations are observed as follows:
\begin{itemize}
\item They have limited coverage of the diverse \varInfoChanShort available to users and, critically, struggle to account for the fluid switching between them (e.g., transitioning from passive media consumption to an active discussion).
\item Their capacity to explain the influence of \varIntVarShort and \varProAttrShort is confined to the specific \gls{isp} for which they were designed. Consequently, they cannot generalize to providers outside their original scope. In other words, they address partial aspects of the complete \frameworkName.
\item Although some models incorporate \varPerVarShort, \varCogVarShort, and \varSitVarShort, these variables are often defined narrowly to fit a specific context. This leaves more general, yet crucial variables -- such as a user's pre-existing beliefs, the process of resolving doubt, or cumulative learning -- under-examined.
\end{itemize}

\section{Application of \frameworkName}
\label{sec:validation}

To demonstrate the utilities of \frameworkName, we applied it to
three pressing issues (misinformation dissemination, authenticity and trust crisis, and dopamine-driven consumption) and discuss its utilities across two dimensions --
characterization and experimental design.

\subsection{Misinformation Dissemination}
\label{sec:validation:misinformation}

\frameworkName allows us to deconstruct the complex misinformation scenario in Figure
\ref{fig:case-study-illustrative-example} not as a single failed search, but as a cascading series of \gls{is} loops, where \emph{many-to-many effect} and
\emph{cross effect} lead to a flawed outcome.

\subsubsection{An \frameworkName Characterization}

The process begins with the information seekers (\name and \othername)
possessing critical \varCogVarShort (they are ``tired of the pandemic'') and
\varPerVarShort (``suspicious''). This emotional state and pre-existing belief make them highly receptive to contrarian information. This combination lowers
the threshold for \actActivatingShort, which in this case is not an explicit
information need, but a latent \emph{gratification need} that confirms
their worldview.

\emph{The Social Media Encounter (Passive Acquisition).}
The first \gls{is} loop starts with \varSitVarShort (\ie library), and
\varInfoChanShort (phone), and \varIntVarShort (social media). The most critical
point here is the information provider (\ie social media), which is optimized for
user engagement, not informational accuracy. When \name and \othername are during
\actAcquiringShort learning from the video's content, the persuasive message
powerfully interacts with their initial \varPerVarShort and causes confirmation bias
(\varCogVarShort).

\emph{The \gls{genai} Reinforcement (Active, Biased Seeking).} The newly
solidified belief and the information gained from gratification further
triggered the need to find evidence. Here, the cross effect is evident that the
new \actInteractingShort is entirely influenced by the previous \gls{is} loop (\ie a biased and leading input question). This incorrect direction of \gls{is}
behavior is strengthened by the provider intent (\varProAttrShort) of
\gls{genai} -- generating content based on the input -- resulting in the
reinforcement of the incorrect belief. The authoritative-sounding answers from a fabricated source (\actTranslatingShort) interacting with beliefs
confirmed by previous \gls{is} loops clear the information seekers' doubt.

\emph{The Corrective Process (Authoritative Intervention).} Here, which
also demonstrates a cross effect, the task was relocated from the
library to a ``Sharing'' seminar (\varSitVarShort), which places \name and
\othername within an influential social setting. The \gls{is} loop began with
\name and \othername acting as information providers during their presentation.
The \teachername, initially served as an information seeker, motivated by the
need to evaluate their response. However, leveraging her domain expertise
(\varPerVarShort), she deemed their findings unconvincing (\actAcquiringShort).
Consequently, her internal state was not updated; instead, she intervened by
directing the information providers (\name and \othername) to provide the source
paper, which activated an information need in \name and \othername, prompting a
role reversal where they became information seekers. The subsequent interaction with authoritative information providers -- the expert and the library database -- was crucial. These sources possess what we term \emph{verifiable authority}
(\ie inherent trustworthiness and the capacity to enable verification). Engaging with these sources prompted the information seekers (\name and \othername) to revise their internal cognitive states and existing knowledge in light of the evidence.

\subsubsection{From Characterization to Intervention: A Rationale for Solutions}

The \frameworkName framework allows us to see how the initial, flawed
\actAcquiringShort in the first \gls{is} loop becomes the ``engine'' of the
misinformation dissemination. When \name and \othername encounter the
information that confirms their existing beliefs, their internal state is
updated. As they switch roles to become information providers, they spread this
misinformation to others who may also be easily activated as them, creating a
cascade of flawed knowledge. 

The \gls{genai} -- \fakeAIname \space -- further worsens this problem as its intent is to generate content regardless of its verifiable status. As you imagine, while cases like \teachername’s are corrected, far more are not, and those uncorrected claims propagate seemingly credible falsehoods that eventually pass as common, yet mistaken arguments.

From a behavioral standpoint, the third \gls{is} loop identifies the solution:
the rumor halts at the  ``wise people'', who are either (1) information seekers
that verify information actively or (2) information providers with
verifiable authority (\eg \teachername). The spread of the misinformation (or
rumor) stops when a wise seeker engages with a trustworthy provider.

From an \gls{ir} system perspective, the characterization not only outlines the
problem but also highlights specific ``failure points''. These serve as crucial
starting points for further research and intervention. Several general research
questions can arise from the reasoning described above. (1) Focusing on the
\varProAttrShort, how can we create a transparent system that informs users of
their provider's intent (\eg social media's engagement and \gls{genai}'s answer
generation)? This approach can help alert seekers' about trusting the
information and verify them explicitly. (2) When targeting the cross effect and
\varIntVarShort (\eg natural language question as queries to \gls{genai}), can the system process queries fairly without disregarding the
seekers' genuine active information needs (\eg \name verifying the effectiveness of using masks on \fakeAIname), especially if they are not seeking ``wisely'' (\eg using the biased queries)?

While there is a broad scope for interventions and research questions, our explicit discussion aims to inspire researchers to characterize \gls{is} and \gls{ir} in the context of misinformation.

\subsection{Authenticity and Trust Crisis}
\label{sec:app:authenticity}

The ``trust crisis'' in AI-generated content is highlighted as a key factor in
\glspl{mie}, and effectively identifying information automatically generated by \glspl{llm} still remains a
challenge~\citep{bevendorff2024overview,bevendorff2025two}. Being an extension of the misinformation issue analyzed in Section~\ref{sec:validation:misinformation}, a
crisis emerges when information seekers repeatedly encounter
authoritative-sounding, yet false or hard-verified, information from \gls{genai}
providers. While a seeker might initially trust the presented information from
that \gls{genai} provider, their trust erodes with each ``failed'' \gls{is} -- that is, each time the information is contradicted by a more
authoritative provider (a credible source, \eg \teachername) or proceed as \name's ``empty search'' (no information found).

Using \frameworkName, we can characterize this as a pattern of negative feedback
that updates the seeker's internal state. Each corrective experience alters the
seeker's \varPerVarShort (\eg their belief in \gls{genai}'s reliability) and
\varCogVarShort (\eg their tiredness of using \gls{genai}). When this pattern
repeats across a population of users and spreads through various
\varInfoChanShort, a systemic distrust of the technology solidifies into a
crisis (\ie common argument about unreliability of \gls{genai}). Therefore, \frameworkName allows us to frame the authenticity and trust crisis not as a single event
but as the cumulative result of flawed \gls{is} loops over time, as concluded in
Section~\ref{sec:validation:misinformation}. To investigate this
issue, we can use \frameworkName to design a user-centric experiment.

\subsubsection{Industry-driven User-Centric Experimental Design.}
Automatic summarization features such as Google's AI Overviews -- which allows users to review information
without necessarily clicking on source articles -- are becoming a widely used \gls{genai} application in web search \cite{reith2024generative}.
This creates a potential trust
challenge, as the system's ability to generate convincing content
(\varProAttrShort) makes external verification difficult. Consider a scenario
where a researcher working at a commercial web search company, \nameforGoogle, is tasked with enhancing user
trust in these summaries.

A key hypothesis is that user trust erodes not just from potentially inaccurate information, but from the difficulty of verifying plausible content (\varProAttrShort). Following \frameworkName, this problem may be improved during the \actAcquiringShort activity, where a user evaluates the quality and relevance of the presented answer. This leads \nameforGoogle to a specific research question: can a simple \gls{ui} intervention -- a credibility indicator (\varIntVarShort) attached to the source reference -- influence a user's trust during the \actAcquiringShort phase?

\frameworkName provides a systematic structure for designing a robust online A/B
test, ensuring that the experiment isolates the intended effect of the \gls{ui} indicator. 

\emph{Controlling Confounding Variables.} The framework serves as a
    checklist to identify and control for other variables that could potentially influence
    user trust. To isolate the effect of the new indicator (\varIntVarShort),
    \nameforGoogle ensures that other factors, such as \varInfoChanShort (\eg indicator
    only performed through laptop access) and \varProAttrShort (\eg no algorithmic
    changes), are kept constant. While \varSitVarShort cannot be directly controlled in an online
    test, effects can be minimized through randomization.

\emph{Defining Evaluation Metrics.} The framework guides the
    measurement of ``trust'' by looking at its effect on a subsequent activity:
    \actActivatingShort. Hypothetically, if the indicator is effective, a user's trust should increase, reducing the need for follow-up verification actions (\ie less unsatisfied need). Therefore,
    \nameforGoogle can measure the change during \actActivatingShort, such as a
    decrease in query reformulations or less attention to search results within and below the AI summary. An improvement in these metrics would suggest that the indicator enhances the user's judgment during \actAcquiringShort.

This structured experimental design allows \nameforGoogle to investigate several
key questions: First, by manipulating the \varIntVarShort, how does it influence subsequent
    \actActivatingShort? Does an indicator of high authoritativeness reduce reactivation by distrust before, while an indicator of low trust increases them? Second, how do these patterns above vary across different user segments,
    as defined by \varPerVarShort, \eg age or technical expertise?
    Third, does the additional \gls{ui} element increase the user's cognitive load
    (\varCogVarShort)? For instance, does it increase the time taken to
    comprehend the summary, and is the intervention's value worth this potential
    cognitive cost?

By applying \frameworkName, \nameforGoogle can move from a
general problem to a comprehensive and methodologically sound experiment. 
\subsection{Dopamine-driven Content Consumption}

The ``dopamine-driven content consumption'' (or \emph{doomscrolling} \cite{sharma2022dark}) represents \glspl{mie} such as social media, where the information providers are optimized to maximize user engagement by exploiting the brain's reward system (\eg dopamine release) \cite{sharpe2025dopaminescrolling}. In this scenario, the provider's primary goal shifts from satisfying an informational need to fostering
compulsive engagement for attention or commercial purposes (\eg advertising or
sales)~\cite{tereshchenko2023neurobiological,westbrook2021striatal}.

From the \gls{ir} system perspective, this phenomenon is best understood as a powerful feedback loop (as described in
Section~\ref{sec:framework:complex-relationships}), where \varProAttrShort directly shape seekers' \actAcquiringShort, \varCogVarShort and subsequent activities (\eg \actActivatingShort), often leading to compulsive use. The process starts with an information provider, whose core attribute is engineered for maximum engagement, not necessarily user well-being or informational accuracy.

This information provider translates (\actTranslatingShort) its vast data repository into presented information (\eg an ``endless scroll'' feed) designed to be endlessly novel and stimulating. Here, a seeker's engagement is often initiated not by a classic knowledge gap, but by a gratification need (\actActivatingShort), such as alleviating boredom. 

During \actAcquiringShort, the seekers' internal state gets altered
(\varCogVarShort) due to, \eg ``dopamine'' -- a feeling of transient pleasure or reward. This reward fulfills the immediate gratification need that initiates the first \gls{is} loop. Crucially, this reward reinforces the belief that the platform is a reliable source of such rewards, \ie the seekers' \varPerVarShort is modified accordingly, and through reinforcement repeatedly. This pattern, \ie dopamine cycle~\citep{de2025social}, keeps delivering the reward for gratification need (\actActivatingShort). In other words, seekers are addicted to the dopamine reward that is triggered by an algorithmically engagement-maximized system, which drives users to stay in continuous, rapid \gls{is} loops that can become compulsive.

\subsubsection{Academic-driven System-Oriented Experiment Design.}
In contrast to Section~\ref{sec:app:authenticity}, we here exemplify how \frameworkName can guide the design of a more controlled, system-centric laboratory experiment. The goal is to design and evaluate a system (information provider) that can mitigate the addictive behavioral patterns described earlier.

Suppose a researcher, \nameforSocialMedia, aims to develop a ``healthier'' social media (\ie \gls{rs}) algorithm. Informed by \frameworkName, they formulate a research question: ``can we alter the \actTranslatingShort logic to generate feeds with ``healthier'' properties (\varProAttrShort) to reduce compulsive engagement loops while maintaining high user satisfaction?''

To investigate, \nameforSocialMedia designs a within-subject lab experiment, which naturally minimizes variance of \varSitVarShort (lab
environment), \varInfoChanShort (same device), \varProAttrShort and
\varCogVarShort (controlled group of participants). Participants interact with
feeds from two different algorithms in counterbalanced sessions, separated by a
time gap to mitigate order and learning effects. This design allows for a direct
comparison of how the manipulated feed properties (\varProAttrShort) affect each
user's engagement behavior (\actInteractingShort). In the \emph{Control Condition}, participants interact with a feed generated by a state-of-the-art engagement-maximization algorithm. The provider's intent (\varProAttrShort) is to prioritize content predicted to generate the most immediate interaction (\ie achieve dopamine hit). In the \emph{Treatment Condition}, participants interact with a feed using an identical \gls{ui} but powered by a modified ``healthier'' algorithm (\varProAttrShort). This new algorithm represents a change to the \actTranslatingShort logic, altering the feed's properties.

The key outcome is to determine if the intervention (\ie any type of healthier
strategy) can achieve high user satisfaction within shorter, less compulsive
engagement periods. Thus, the evaluation is designed based on the activity --
\actAcquiringShort and \actActivatingShort, such as (1) overall satisfaction
through a questionnaire where users rate the relevance, and overall quality of
the content they consumed; (2) engagement session duration, which approximately represents the time users holding the information need (\ie when users stop
engaging, they are considered as satisfied and stop triggering dopamine hit);
(3) system-level properties, such as content diversity scores. 

By investigating the manipulation of \varProAttrShort in this controlled manner, this \frameworkName-informed approach pinpoints a lever for improving the information ecosystem, paving the way for more responsible and ethical system design.

\section{Conclusions and Future Work}
\label{sec:conclusion}

Contemporary information access operates within an increasingly complex ecosystem of technologies, user behaviors, and contexts. While it is easy to envision \gls{is} scenarios spanning multiple devices, modalities (e.g., text and audio), and systems (e.g., search engine and \gls{genai}-powered chat), it remains challenging to characterize \glspl{mie} with an \gls{is} model that captures this full complexity. We introduce the Information Seeking in Modern Information Environments (\frameworkName) framework as a first step to provide vocabularies and concepts. With \frameworkName, we analyze six \gls{is} and \gls{ir} models, highlighting the need for novel alternatives to characterize \glspl{mie}.
We further applied \emph{\frameworkName} to three pressing issues (misinformation, authenticity and trust crisis, and dopamine-driven consumption) to demonstrate its utility in formulating research questions and guiding experimental design.

\emph{Limitations.} While our \frameworkName framework is designed to cover core \emph{Components}, \emph{Variables}, and \emph{Activities} considered in \glspl{mie}, its broader positioning would benefit from co-design with multidisciplinary teams, particularly across adjacent fields such as HCI and cognitive science.
Another limitation is the framework's abstract nature: it does not, by itself, address privacy and fairness. These concerns are especially acute in applications that require large-scale user data. A critical next step is to embed privacy-preserving techniques and fairness-aware metrics when operationalizing \frameworkName. A full discussion is beyond the scope of this paper, but promising directions include privacy-preserving personalization via federated learning~\cite{FabioFederated2023} and equitable outcomes via fairness-aware ranking~\cite{Sahin2019fairnessranking}. Applying \frameworkName in concert with such practices would strengthen its ethical foundations.

\emph{Future Work.} \frameworkName is presented as a conceptual and analytical tool. Future work is needed to operationalize its high-level concepts into more practical design principles and concrete evaluation methodologies. \frameworkName can be leveraged to bridge the gap between qualitative, user-centric studies and offline, system-oriented evaluations.
We also leave to future work an empirical validation of our \frameworkName framework, which involves designing and conducting a series of observational studies and controlled experiments across diverse user populations, tasks, and technological contexts -- including multiple devices, modalities, and systems such as conversational search and social media.

\begin{acks}
This research was conducted by the ARC Centre of Excellence for Automated Decision-Making and Society (ADM+S, CE200100005), and funded fully by the Australian Government through the Australian Research Council. This work was conducted on the unceded lands of the  Woi wurrung and Boon wurrung language groups of the eastern Kulin Nation. We pay our respect to Ancestors and Elders, past and present, and extend that respect to all Aboriginal and Torres Strait Islander peoples today and their connections to land, sea, sky, and community. 
\end{acks}

\clearpage


\bibliographystyle{ACM-Reference-Format}
\balance
\bibliography{ref}

\end{document}